\documentclass[final            % use final for the camera ready runs
  ]
  {aipproc}

\layoutstyle{8x11single}

\begin{document}

\title{Overview of Dynamic Models for $ \gamma N \rightarrow \Delta$ }


%%\author{<D. Drechsel>}{
%% address={<author1 address>}
%%}

\author{D. Drechsel and L. Tiator}{
  address={Institut für Kernphysik, Universität Mainz, 55099 Mainz, Germany}
}

%%\author{<>}{
%% address={<common address for author2 and author3>}
%% ,altaddress={<author1 address>} % additional visiting address
%%}

\begin{abstract}
 We present an overview of dynamic and related models for pion
 electroproduction on the nucleon due to $\Delta$(1232) excitation. The
 electromagnetic multipoles and their ratios obtained from these models and
 from effective field theories are compared, and the existing contradictory
 results for the pion cloud effects are discussed.
\end{abstract}

\maketitle

%%%%%%%%%%%%%%%%%%%%%%%%%%%%%%%%%%%%%%%%%%%%
%% MAINMATTER
%%%%%%%%%%%%%%%%%%%%%%%%%%%%%%%%%%%%%%%%%%%%

\section{Introduction}

The $\Delta(1232)$ or $P_{33}$ is the strongest resonance in the pion ($\pi$)
and photon ($\gamma$) induced reactions on the nucleon (N). It is the only
isolated resonance of the nucleon, and the Argand diagram of $\pi$-N scattering
is a perfect circle in the $\Delta(1232)$ region up to c.m. energies $W \approx
1400$~MeV. In the framework of a quark model with SU(6) symmetry, the
$\Delta(1232)$ and the nucleon constitute a degenerate ground state of the
3-quark system. This symmetry is broken by a spin-dependent "hyperfine"
interaction due to a residual force of the quark-gluon interaction. Because of
its spin dependence this interaction also leads to a tensor force, which yields
non-spherical (D state) components in the wave functions. Whereas the
photoexcitation process $\gamma$ + N $\rightarrow \Delta(1232)$ is dominated by
a magnetic dipole transition (M1), these D states give rise to a small electric
quadrupole transition (E2). However, the experimental ratio $R_{EM}$ = E2/M1$
\approx -2.5~\%$ can only be reached with an unnaturally large bag radius of
about 1 fm. It is therefore now generally assumed that a large contribution of
the E2 value stems from the pion cloud dominating the large-distance physics.
\\

The aspects of short- and long-distance physics were first studied in chiral
bag models by describing the nucleon with an interior quark bag of radius $r_0
\approx 0.5$ fm and a pion cloud extending to radii $r \approx 1.5$ fm. Since
the quadrupole moment scales with $r^2$, the break-down in pion cloud and bag
contributions depends strongly on the chosen radius $r_0$, although the overall
size of $R_{EM}$ is relatively stable and in reasonable agreement with the
data. Nowadays the long-range physics is systematically described by effective
field theories (EFTs) based on an expansion in the external momenta, the pion
mass, and the N-$\Delta$ mass splitting. The effects of the short-distance
physics are absorbed in low-energy constants (LECs), defined by the
coefficients in front of the most general lagrangians in accordance with QCD.
However, the splitting in short- and long-distance physics is still not unique,
because it depends on the scale $\lambda$ used in the renormalization process
of the loop integrals.
\\

Another, more phenomenological approach to pion scattering and pion
photoproduction is given by dynamic models (DMs). Whereas EFTs are an exact
representation of QCD in the sense that they produce a perturbation series
based on the most general lagrangian with the symmetries of QCD, a DM starts
from a simplified but reasonable lagrangian that is treated to all orders by
means of  Lippman-Schwinger or similar equations. As a result the two
approaches are complementary. Dynamic models nicely describe the $\pi$-N
scattering in the resonance region and thus yield a unitary scattering
amplitude, whereas they have problems with gauge invariance and relativity. On
the other hand, EFTs are manifestly Lorentz and gauge invariant, whereas the
perturbative approach can not be expected to give a fully unitary amplitude.
\\

Because of its importance in pion scattering and pion photoproduction, the
$\Delta$ is responsible for several ground state properties of the nucleon by
sum rules:
\begin{itemize}
\item
The Adler-Weisberger sum rule connects the axial coupling constant $g_A$ with
an integral over $\sigma_{tot}^{\pi ^+ p} -\sigma_{tot}^{\pi ^- p }$, the
difference of the total cross sections for charged pion scattering off the
proton. The $\Delta(1232)$ turns out to be essential to saturate this sum rule.
\item
The Gerasimov-Drell-Hearn sum rule expresses the square of the anomalous
magnetic moment, $\kappa_N^2$,  by an integral over $\sigma_{3/2}
-\sigma_{1/2}$, the difference of the helicity dependent total cross sections
for photoabsorption on the nucleon. The integral over the $\Delta(1232)$ region
yields the sum rule value within about 10~\%.
\item
According to Fubini, Furlan, and Rossetti, the Pauli form factor $F_2^N(Q^2)$
is proportional to an integral over the first relativistic  structure function
for neutral pion electroproduction, $A_1^{(+,0)}$. Again, the integral over the
$\Delta(1232)$ region contributes about 90~\% to the sum rule value, the
remainder being due to dipole absorption near threshold and in the second
resonance region.
\item
Related to the low energy theorem of Nambu, Lourié, and Shrauner, a further
theorem of Fubini, Furlan, and Rossetti gives a connection between
$F_1^V(Q^2)-G_A^V(Q^2)$, the difference of the isovector Dirac and axial form
factors, and an integral over the sixth relativistic  structure function
$A_6^{(-)}$ for charged pion electroproduction. The relative closeness of the
two radii is expressed by the fact that the large positive contribution of the
$\Delta(1232)$ is essentially canceled by large negative contributions mainly
from the second resonance region.
\end{itemize}

In conclusion the structure of the nucleon is intimately interwoven with the
properties of the $\Delta(1232)$. This also applies to the question whether
elementary particles like the nucleon are "intrinsically" spherical or deformed
objects. In fact in a simple quark model, the hyperfine tensor force will
inevitably admix D-state components to both nucleon and $\Delta(1232)$ wave
functions. In particular, the "deformation" leads to a quadrupole component in
the N-$\Delta(1232)$ transition, which will be discussed in the following
sections. In principle one could also think about the "reorientation effect"
$\Delta(1232) \rightarrow \gamma + \Delta(1232)$ appearing as an intermediate
state in the reaction $\gamma + N \rightarrow \gamma \, ' + \pi^0 + N \, '$.
This reaction  is presently under investigation at MAMI with the aim to
determine the magnetic dipole moment of the $\Delta(1232)$, which should take a
complex and energy-dependent value  due to the finite width of the
$\Delta(1232)$. However, the electric quadrupole moment has no diagonal value
because of parity and time-reversal invariance. It is therefore unlikely that
radiative pion photoproduction will ever teach us something about the
deformation of the $\Delta(1232)$ proper, although we expect a sizeable oblate
quadrupole moment of the $\Delta(1232)$.

\section{Basics}

\subsection{Scattering matrix}

Let us recall some basic definitions for the simple case of pion-nucleon
scattering below the two-pion threshold and with the approximation that the
electromagnetic interaction can be neglected. Because of angular momentum and
parity conservation for the strong interaction, the S-matrix is then diagonal
in a representation with quantum numbers $l$ (orbital angular momentum), $J=l
\pm 1/2$ (total angular momentum), and isospin $I=1/2$ or $3/2$. All the
physics is contained in the scattering phases $\delta_{l \pm}^I(W)$, where W is
the total energy in the c.m. frame. (Note that we suppress the isospin index
$I$ in most of the following text.) The S-matrix elements are
\begin{equation}
S_{l \pm}(W)=e^{2 i  \delta_{l \pm}(W)}. \label{eq:2.1}
\end{equation}
The S-matrix is unitary, i.e., $ S^\dagger (W) S(W) =1$ and  $S^*_{l
\pm}(W)S_{l \pm}(W)=1$. Next we define the elements of the T- and K-matrices by
\begin{equation}
S_{l \pm}(W)=1 + 2iT_{l \pm}(W)= \frac{1+iK_{l \pm} (W)}{1-iK_{l \pm}(W)}\,.
\label{eq:2.2}
\end{equation}
These matrix elements can be expressed by the scattering phases,
\begin{equation}
T_{l \pm}(W)= \frac{K_{l \pm} (W)}{1-iK_{l \pm}(W)}= \mbox{sin} \delta_{l
\pm}(W)e^{i \delta_{l \pm}(W)}\,, \label{eq:2.3}
\end{equation}
\begin{equation}
K_{l \pm}(W)= \frac{T_{l \pm} (W)}{1+iT_{l \pm}(W)}= \mbox{tan} \delta_{l
\pm}(W)\,. \label{eq:2.4}
\end{equation}
According to these definitions, the T-matrix is purely imaginary at resonance,
$\delta_{l \pm}(W_R)=\pi /2$, whereas the K-matrix has a pole at this point. In
particular the $\Delta(1232)$ resonance with quantum numbers $l=1, J=3/2$ leads
to a pole, $K_{1+}(W) \rightarrow \infty$ if $W \rightarrow 1232$~MeV, whereas
$T_{1+}=i$ in this limit. On the other hand, a resonance is defined by a pole
of the S-matrix (or T-matrix) in the complex energy plane, with negative
imaginary part of the energy. In particular, the $\Delta(1232)$ pole lies at
$W_R \approx (1210-50i)$~MeV. In case of approximations, which are inevitable
in most calculations, it is recommended to approximate the K-matrix elements.
As shown by Eq.~(\ref{eq:2.2}), the S-matrix is unitary for any real value of
$K=V$, whereas $T=V$ results in $ S^\dagger S =1+4V^2$ and violates unitarity
at second order in the potential.

\subsection{Resonances and speed plot}

The analytic continuation of a resonant partial wave as function of energy
should generally lead to a pole in the lower half-plane of the second Riemann
sheet. A pronounced narrow peak reflects a time-delay in the scattering process
due to an unstable excited state. This time-delay is related to the speed of
the scattering amplitude~\cite{Hoe92, Han96},
\begin{equation}
SP(W)= \big|\frac{dT(W)}{dW} \big|\, . \label{eq:2.4a}
\end{equation}
In the vicinity of the resonance pole, the energy dependence of the full
amplitude, $T(W)=t(W)+T_R(W)$, is given by the resonance contribution,
\begin{equation}
T_R(W) \rightarrow \frac{Z \; \Gamma_R }{M_R-W-i \Gamma_R/2} \, ,
\label{eq:2.4b}
\end{equation}
whereas the background contribution $t(W)$ should be a smooth function, ideally
a constant. The speed has its maximum at $W=M_R$, $SP(M_R)=H$, and the
half-maximum values are $SP(M_R \pm \Gamma_R /2)=H/2$. This yields the pole
position of the resonance at $W=M_R - i \Gamma_R /2$. The residue of the pole
is determined by the resonance pole parameter Z, which is a complex number with
absolute value r and phase $\Phi$. The latter is due to the interference with
the background, it may be determined from the phase of the complex speed vector
$dT/dW$ at the pole. The Particle Data Group~\cite{Yao06} gives the following
residue for pion-nucleon scattering at the position of the $\Delta (1232)$: $r
\approx (53 \pm 3)$~MeV$/\Gamma_R$ and $\Phi \approx (- 47 \pm 1 )^{\circ}$. In
the case of pion photoproduction, the  strong interaction takes place only
after the resonance excitation, and therefore we may expect a phase rotation of
about $-20^{\circ}$ to $-25^{\circ}$ by the interference between background and
resonance.

\subsection{Electromagnetic multipoles}

We consider the reaction $\gamma^*(k, \epsilon )+ N(p,s) \rightarrow  \pi (q)+
N'(p',s')$, with $k, p, q$, and $p \, '$ the momenta of the respective
particles, $\epsilon$ the polarization of the virtual photon $\gamma^*$, and
$s$ and $s'$ the polarization of the nucleon in the initial and final state,
respectively. The matrix elements of the transition four-current ($\rho,\vec
J$) are classified as transverse electric (E), transverse magnetic (M),
longitudinal (L), and Coulomb or "scalar" currents (S). Denoting the isospin
with $I$, a pion-nucleon state can be excited by 4 multipole transitions,
\begin{equation}
{\mathcal M}_{l \pm}^I (W, Q^2)=
(E_{l \pm}^I, M_{l \pm}^I, L_{l \pm}^I, S_{l
\pm}^I)\,, \label{eq:2.5}
\end{equation}
with the exception that the multipoles  $M_{0 \pm }^I, E_{0-}^I, E_{1-}^I,
L_{0-}$ and $S_{0-}$ do not exist. The longitudinal and Coulomb multipoles are
related by gauge invariance, $\vec k \cdot \vec J = \omega \rho$, which leads
to
\begin{equation}
 |\vec k| \, L_{l \pm}^I (W, Q^2)= \omega \, S_{l \pm}^I (W, Q^2)\,. \label{eq:2.6}
\end{equation}
Since the photon c.m. energy  $\omega$ vanishes for $Q^2=Q^2_0=W^2-M^2$, the
longitudinal multipole must have a zero at that momentum transfer, $L_{l
\pm}^I(W,Q^2_0)=0$. Furthermore, the longitudinal and Coulomb multipoles have
to become equal in the real photon limit, $L_{l \pm}^I(W,Q^2=0)=S_{l
\pm}^I(W,Q^2=0)$. Another important property is the model-independent behavior
of the multipoles at physical threshold (vanishing pion momentum, $q= |\vec
q|=0)$ and at pseudothreshold (Siegert limit, vanishing photon momentum $k=
|\vec k|=0)$:
\begin{eqnarray}
(E_{l+}^I, L_{l+}^I) &\rightarrow& k^l q^l \;\; (l \ge 0)\, \nonumber\\
(M_{l+}^I, M_{l-}^I) &\rightarrow& k^l q^l \;\; (l \ge 1)\, \nonumber\\
(L_{1-}^I) &\rightarrow& kq \;  \nonumber\\
(E_{l-}^I, L_{l-}^I) &\rightarrow& k^{l-2} q^l\;\; (l \ge 2)\ \,.
\label{eq:2.7}
\end{eqnarray}
According to Eq.~(\ref{eq:2.6}) the Coulomb amplitudes acquire an additional
factor $k$ at pseudothreshold, i.e., $S_{l \pm}^I \sim k L_{l \pm}^I$. This
limit is reached at $Q^2=-(W-M)^2$, and because no direction is defined at
$\vec k = 0$, the electric and longitudinal multipoles have to be related at
this point,
\begin{equation}
 E_{l+}^I/L_{l+}^I \rightarrow 1 \;\;  \mbox{and} \;\; E_{l-}^I/ L_{l-}^I \rightarrow -l/(l-1)
 \;\;  \mbox{if} \;\;  k \rightarrow 0. \label{eq:2.8}
\end{equation}

\subsection{Fermi-Watson theorem}

Below the two-pion threshold the excited nucleon can only decay by emitting a
photon or a pion. The S-matrix then takes the general unitary form:
\begin{equation}
S = \left( \begin{array} {cc}
S_{\pi \pi} & S_{\gamma \pi} \\
S_{\pi \gamma} & S_{\gamma \gamma}
\end{array} \right)
= \left( \begin{array} {cc}
\sqrt{1-\eta ^2} \; e^{2 i \delta _{\pi N}} & i \eta \, e^{ i (\delta _{\pi N} + \delta _{\gamma N})} \\
i \eta \, e^{ i (\delta _{\pi N} + \delta _{\gamma N})} & \sqrt{1-\eta ^2} \;
e^{2 i \delta _{\gamma N}} \;
\end{array} \right) \, ,
\label{eq:2.9}
\end{equation}
where the real number $\eta$ is the inelasticity, and $\delta _{\pi N}$ and $
\delta _{\gamma N}$ are the pion-nucleon and  Compton scattering phases,
respectively. Since the electromagnetic interaction is small, we may neglect
the terms of order $\eta^2$. These approximations lead to the following
T-matrix elements: $T_{\pi \pi}=\mbox{sin} \; \delta _{\pi N}e^{i \delta _{\pi
N}}$, $T_{\gamma \pi}=\pm e^{i \delta _{\pi N}} \; |T_{\gamma \pi}|$, and
$T_{\gamma\gamma}=0$. Because the transition matrix element $T_{\gamma \pi}$ is
proportional to the electromagnetic multipoles of Eq.~(\ref{eq:2.5}),
$T_{\gamma \pi}= \sqrt{2 \, q(W) \, k(W,Q^2)} \, {\mathcal M}$, also the
electromagnetic multipoles carry the phase of the respective pion-nucleon
states, for example, ${\mathcal M}_{1+ }^{3/2} (W, Q^2)= \pm \, e^{i
\delta_{1+}^{3/2} (W)} \; |{\mathcal M}_{1+ }^{3/2} (W, Q^2)|.$

\subsection{Form factors}

We restrict the following discussion to the $\Delta (1232)$ resonance.
Generalizations to higher resonances have been given, but the definitions
become more involved. Corresponding to the independent transition multipoles,
the following three form factors for the N-$\Delta$ transition can be defined:
\begin{eqnarray}
M_{1+}^{3/2} (M_\Delta , Q^2) &=&i \, N \; \frac{k_\Delta (Q^2)}{M_N} \, G_M^* (Q^2) \, \nonumber\\
E_{1+}^{3/2} (M_\Delta , Q^2) &=& -i \,  N \; \frac{k_\Delta (Q^2)}{M_N} \, G_E^* (Q^2) \, \nonumber\\
S_{1+}^{3/2} (M_\Delta , Q^2) &=& -i \,  N  \; \frac{k_\Delta (Q^2)^2}{2 \, M_N
M_\Delta} \, G_C^* (Q^2) \, , \label{eq:2.10}
\end{eqnarray}
with $N=\sqrt{3 \alpha /(8 q _\Delta \Gamma_\Delta)}$, where $\alpha \approx
1/137$, and the photon ($k_\Delta $) and pion ($q_\Delta$) three-momenta have
been evaluated at the resonance. The proportionality factor contains kinematics
that relates pion photoproduction to total photoabsorption at the resonance.
Comparing with Eq.~(\ref{eq:2.7}) we note that these definitions divide out the
$k$ dependence at pseudothreshold such that the form factors are finite at this
point. Equation~(\ref{eq:2.10}) corresponds to the definition of Ash~
\cite{Ash67}, the form factors of Jones and Scadron~\cite{Jon73} are obtained
by multiplication with an additional factor, $G^{JS}=\sqrt{1+Q^2/(M_N
+M_\Delta)^2} \, G^{Ash}$.

\section{Dynamic models}

A dynamic model in the $\Delta(1232)$ region contains the Hilbert vectors $|N
\pi>,  |N \gamma>$, and $|\Delta(1232)>$. The first step is to calculate the
pion-nucleon scattering states, in a second step we include the electromagnetic
interaction of the real or virtual photons with the pions and nucleons. A good
introduction to this topic is given by Ref.~\cite{Bur04}, which also includes
references to earlier work and a useful summary on various multi-channel
K-matrix models that are needed to describe the higher energy region with many
open channels.

\subsection{The pion-nucleon system}

Dynamic models treat the hadronic system on the basis of a free hamiltonian, a
quasi-potential, and "bare" transition operators describing the excitation of
resonances. This hamiltonian is treated in the framework of a Bethe-Salpeter
equation with the result of a fully unitary T-matrix, see, for example,
Refs.~\cite{Sat96, Sat01}. However, the quasi-potential does not allow the
(virtual) production of heavier mass systems, for example, two-pion
intermediate states are excluded. As a consequence, the model is not truly
relativistic, even if the kinematics is relativistic. The quasi-potential
v$(\pi N \rightarrow \pi N)=$v$_{\pi N}$ describes a non-resonant background
and is typically given by s- and u-channel Born terms and t-channel meson
exchange. The $t$-matrix for the background takes the form
\begin{equation}
 t_{\pi N} (W)=  \mbox{v} _{\pi N} [ 1 + g_{\pi N} (W)\, t_{\pi N} (W)] \, , \label{eq:3.1}
\end{equation}
where $g_{\pi N}$ is the propagator of the free pion-nucleon system. The
background modifies the "bare" transition operator $\Gamma_{\pi N \rightarrow
\Delta}$, which leads to the "dressed" $ N \Delta$ vertex
\begin{equation}
\tilde\Gamma_{\pi N \rightarrow \Delta}(W) =\Gamma_{\pi N \rightarrow
\Delta}[1+g_{\pi N} (W) \, t_{\pi N }(W)] \,, \label{eq:3.2}
\end{equation}
as well as the free $\Delta$ propagator $G_\Delta$, which builds up the
"dressed" $ N \Delta$ propagator
\begin{equation}
\tilde{G_\Delta} (W)=G_\Delta (W) [1+\Sigma_\Delta (W) \,\tilde{G_\Delta} (W)]
=[W-M_\Delta -\Sigma_\Delta (W)]^{-1} \,,\label{eq:3.3}
\end{equation}
with $\Sigma_\Delta$ the complex self-energy due to intermediate pion-nucleon
loops. With these definitions the full T-matrix for pion-nucleon scattering
takes the form
\begin{equation}
T_{\pi N}(W)= t_{\pi N}(W) + \tilde\Gamma_{\Delta\rightarrow \pi N}(W) \,
\tilde G_\Delta(W)\, \tilde\Gamma_{\pi N \rightarrow \Delta}(W)\,.
\label{eq:3.4}
\end{equation}
This Bethe-Salpeter equation is solved for each multipole ($l \pm$) and isospin
$(I)$ channel, and the parameters contained in both quasi-potential and bare
$\pi N \rightarrow \Delta$ vertex are fitted to the experimental pion-nucleon
phase shifts $\delta_{l \pm}^I (W)$.

\subsection{The electromagnetic interaction}

If the photon is absorbed by the target nucleon, it will either produce a
pion-nucleon state by the transition potential v$_{\gamma N \rightarrow \pi N}$
or a resonance by the bare vertex $\Gamma_{\gamma N \rightarrow \Delta}$. The
electromagnetic background radiation is described by
\begin{equation}
t_{\gamma N \rightarrow \pi N} (W,Q^2)= \mbox{v} _{\gamma N \rightarrow \pi
N}(W, Q^2) + t_{\pi N} (W) \, g_{\pi N}(W)\, \mbox{v} _{\gamma N \rightarrow
\pi N}(W, Q^2)\,. \label{eq:3.5}
\end{equation}
The second term on the r.h.s of this equation describes the rescattering, to
lowest order a pion-nucleon loop. The electromagnetic interaction excites the
hadronic system to a virtual state with energy $W'$, which propagates with
$g_{\pi N}(W)=(W-W'+i \epsilon)^{-1}=  - i \pi \delta (W-W') + {\mathcal
P}^(W-W')^{-1}$. This splits the rescattering term in on-shell and off-shell
contributions, the latter ones being described by a principal value integral
over all pion-nucleon scattering states with energy $W'$. A cautionary remark:
In the work of Sato and Lee~\cite{Sat96, Sat01} the rescattering in
Eq.~(\ref{eq:3.5}) is due to the non-resonant pion-nucleon interaction, that
is, with the resonance switched off. In contrast, the DMT model~\cite{Kam99,
Kam01} uses the full pion-nucleon interaction in this equation.
\\

The potential v$_{\gamma N \rightarrow \pi N}$ is given by the s- and u-channel
pion-nucleon Born terms, the Kroll-Ruderman term, and t-channel contributions
from the exchange of neutral pions or vector mesons ($\rho, \omega$). The
meson-exchange diagrams contain some freedom, because the coupling between the
mesons and the nucleon is not too well known. A further problem arises for the
coupling of virtual photons. In principle, the s- and u-channel terms should
appear with the Dirac and Pauli form factors of the nucleon, the Kroll-Ruderman
term with the axial form factor of the nucleon, and the t-channel terms with
pion or pion-meson transition form factors. However, a simple multiplication
with the respective form factors violates gauge invariance. Instead, a
consistent description requires that the form factors be built up from a gauge
invariant many-body hamiltonian of the hadronic system, for example, in ChPT by
loop diagrams to a given order of expansion. Short of a complete theory, one
can either set all form factors equal, for example, to the familiar dipole
form, or enforce gauge invariance explicitly by the replacement $J_\mu
\rightarrow J_\mu- k_\mu \; J \cdot k /k\cdot k$. Although the latter procedure
ensures gauge invariance, it is by no means unique. The bare vertex for the
reaction $\gamma + N \rightarrow \Delta$  can be expressed by 3 transition form
factors $G_{M1}(Q^2), G_{E2}(Q^2)$, and $G_{C2}(Q^2)$. The vertex becomes
dressed by inserting all rescattering "bubbles",
\begin{equation}
\tilde\Gamma_{\gamma N \rightarrow \Delta} (W,Q^2)=\Gamma_{\gamma N \rightarrow
\Delta} (Q^2) +\tilde\Gamma_{\pi N \rightarrow \Delta} (W,Q^2)  \, g_{\pi N}(W)
\, \mbox{v} _{\gamma \pi}(Q^2) \, . \label{eq:3.5a}
\end{equation}
Combining these results we obtain the full $T$-matrix for the reaction $\gamma+
N \rightarrow \pi + N \,'$,
\begin{equation}
T_{\gamma N \rightarrow \pi N}(W,Q^2)= t_{\gamma N \rightarrow \pi N}(W,Q^2) +
\tilde\Gamma_{\Delta\rightarrow \pi N}(W) \,\frac{1}{W-M_\Delta-\Sigma_\Delta
(W)} \, \tilde\Gamma_{\gamma N \rightarrow \Delta}(W,Q^2)\,. \label{eq:3.5b}
\end{equation}
This is again the definition of Sato and Lee. In order to compensate for the
different form of the background, the dressed vertex $\tilde\Gamma_{\gamma N
\rightarrow \Delta}$ in Eq.~(\ref{eq:3.5b}) has to be replaced by the bare
vertex $\Gamma_{\gamma N \rightarrow \Delta}$ in the DMT model.
\\

We repeat that pion rescattering is treated by summing up all diagrams with an
arbitrary number of pion-nucleon bubbles in the intermediate states. However,
these loops require a phenomenological cut-off $\Lambda_{\pi N}$ to regularize
the integrals for large off-shell momenta. We note that the rescattering
effects are contained in all of the functions on the r.h.s. of
Eq.~(\ref{eq:3.5b}). Furthermore, the rescattering process will contain
model-dependent off-shell effects, even if all the pion-nucleon phase shifts
are correctly reproduced. Let us now summarize the model-dependencies and
shortcomings of the state-of-the-art dynamic models:
\begin{itemize}
\item
the choice of the quasi-potential v$_{\pi N}$ of pion-nucleon scattering and
the cut-off $\Lambda_{\pi N}$ for regularization of the loop integrals,
\item
the coupling constants of the vector mesons to pions and nucleons,
\item
the  bare vertices $\Gamma_{\pi N \rightarrow \Delta}$ and $\Gamma_{\gamma N
\rightarrow \Delta}$ for $\Delta (1232)$ production by pions and photons,
respectively,
\item
the choice of the bare N$\Delta (1232)$ transition form factors $G_{M1}(Q^2),
G_{E2}(Q^2)$, and $G_{C2}(Q^2)$ corresponding to the multipoles $M_{1+},
E_{1+}$, and $S_{1+}$, respectively,
\item
the neglect of intermediate states other than pion-nucleon loops, and of such
loops before the photon has been absorbed, and
\item
the treatment of backgrounds from u-channel $\Delta$ excitation and higher
resonances.
\end{itemize}

\subsection{Comparison of dynamic and other phenomenological models}

\subsubsection{Sato-Lee model}

Sato and Lee (S-L) derive their dynamic model from a lagrangian in all parts
~\cite{Sat96, Sat01}, in particular also the bare vertex for the $\gamma + N
\rightarrow \Delta$ transition is based on an effective lagrangian. The
electromagnetic background is constructed according to Eqs.~(\ref{eq:3.1}) and
(\ref{eq:3.5}), that is, the pion produced by the non-resonant electromagnetic
background is rescattering with the non-resonant part of the pion-nucleon
interaction. Therefore the background amplitude carries the slowly varying
phase $\delta_b (W)$ of the pion-nucleon background. The exact value of this
phase is certainly somewhat model-dependent. As an example, in the work of
Höhler~\cite{Hoe92} this phase slowly decreases from $-19.5 ^{\circ}$ at pion
production threshold to about $-24^{\circ}$ at $W=1232$~MeV. The cut-off for
the $\pi N$ loop integrals has a dipole form, $F(\vec{q})=(1+\vec{q} \,
^2/\Lambda^2)^{-2}$ with $\Lambda $ in the range of (690$\pm$50)~MeV, which
corresponds to an r.m.s radius of nearly 1 fm. If the $\Delta$~(1232) is
produced or annihilated, a form factor of similar range is applied. The
regularization procedure is fixed by the strong interaction, no additional
regularization factors are introduced in pion electroproduction.

\subsubsection{DMT model}

The Dubna-Mainz-Taipei (DMT) dynamic model differs from the S-L model by
parameterizing the $\Delta$ contribution to the reaction $\gamma N \rightarrow
\pi N$ right away in the form~\cite{Kam99, Kam01}
\begin{equation}
t_{\gamma N \rightarrow \pi N}^{\Delta , \lambda}(W,Q^2)= {\cal A}^{\Delta ,
\lambda} (Q^2) \, f_{\pi \Delta}(W) \,\frac{\Gamma_\Delta M_\Delta }
{M_\Delta^2-W^2-i M_\Delta \Gamma_\Delta} \, f_{\gamma \Delta}^\lambda (W) \,
e^{i \Phi _\Delta (W)} \, , \label{eq:3.6}
\end{equation}
where $\Phi_\Delta (W)$ corrects the phase of the Breit-Wigner form to the
experimental pion-nucleon phase shift $\delta _{33}(W)= \delta _{1+}^{3/2}(W)$
and $\lambda$ stands for the electromagnetic multipolarities E, M, or S. The
$Q^2$ dependence of the transverse multipoles is parameterized as
\begin{equation}
{\cal A}^{\Delta , \lambda} (Q^2) = {\cal A}^{\Delta ,\lambda} (0) \,
\frac{k_\Delta (Q^2)}{k_\Delta (0)} (1+ \beta Q^2)e^{-\gamma Q^2} G_D (Q^2) \,
, \label{eq:3.6a}
\end{equation}
where $G_D (Q^2)= (1+Q^2/0.71 \mbox{GeV} ^2)^{-2}$ is the dipole form factor,
$\beta$ and $\gamma$ are free parameters to be determined from the data, and
the explicit factor $k$ in front provides the correct behavior at
pseudothreshold.
\\

At variance with the S-L model, the DMT model describes both resonant and
non-resonant electromagnetic amplitudes with the full pion-nucleon interaction,
that is, both contributions carry the full phase $\delta _{33}(W)$. The
convergence of the loop integrals is enforced by a dipole cut-off $F(\vec{q} \,
',\vec{q})=[(\Lambda^2+\vec{q} \, ^2) / (\Lambda^2+\vec{q} \, ' \, ^2)]^2$ at
the vertex leading from an on-shell pion with momentum $\vec{q}$ to an
off-shell pion with momentum $\vec{q} \; '$, and the cut-off is $\Lambda =
440$~MeV.

\subsubsection{MAID model}

The MAID model~\cite{Dre99} uses the pion-nucleon scattering phases as input.
The resonance contribution to pion electroproduction is parameterized as in
Eq.~(\ref{eq:3.6}), and the non-resonant term is treated in an on-shell
approximation, i.e., the principle value integral of the loop contribution is
dropped. The resulting background contribution to the reaction $\gamma N
\rightarrow \pi N$ takes the form
\begin{equation}
t_{\gamma N \rightarrow \pi N}^{bg, \alpha} (W,Q^2) \rightarrow (1+i T_{\pi N
\rightarrow \pi N }^{\alpha}) \; K_{\gamma N \rightarrow \pi N }^\alpha
\rightarrow e^{i \delta_{\alpha} (W)} \, \mbox{cos} \delta_{\alpha} (W) \,
\mbox{v} _{\gamma N \rightarrow \pi N }^\alpha (W,Q^2) \, , \label{eq:3.7}
\end{equation}
where v$_{\gamma N \rightarrow \pi N }^\alpha$ is the quasi-potential for the
non-resonant background terms, and $\alpha$ denotes the quantum numbers. The
background includes the usual nucleon Born terms as well as pion and vector
meson t-channel contributions. The pion-nucleon scattering is described by a
mixture of pseudovector (pv) and pseudoscalar (ps) couplings, such that the
threshold scattering is purely pv as required by chiral invariance, whereas
with increasing energy the interaction becomes more and more ps. The reason for
this parameterization is that the additional gradient in pv coupling turns into
a pion momentum and thus leads to an unphysical increase of the interaction for
high energies, which in other models is damped by momentum-dependent form
factors. We further point out that the ps and pv Born terms differ only in
channels with total angular momentum $J=1/2$, i.e., S wave (0+) and Roper (1-)
multipoles.

\subsubsection{Yerevan-JLab model}

The ps-pv mixing in MAID was introduced in order to avoid cut-off form factors.
As was pointed out by the Yerevan-JLab group~\cite{Azn03}, the original MAID
model still showed a large increase of the Born terms at higher energies, which
had to be compensated by unrealistically large resonance tails. It was
therefore suggested to improve the MAID background by a gradual transition to
the Regge behavior as function of $s=W^2$,
\begin{equation}
t_{YL}^{bg, \alpha} (W,Q^2)= \frac {t_{MAID}^{bg, \alpha} (W,Q^2)}{1+(s/s_{thr}
-1)^2}+ \frac {(s/s_{thr} -1)^2 \, t_{Regge}^{bg, \alpha} (W,Q^2)}{1+(s/s_{thr}
-1)^2} \, , \label{eq:3.8}
\end{equation}
with a Regge background containing the t-channel exchange of $\pi ,\rho ,
\omega , b_1$, and $a_2$ mesons. As is evident from Eq.~(\ref{eq:3.8}), this
description takes exactly the MAID value  at threshold (pv coupling), but
reduces the MAID background (a mixture of ps and pv coupling) to about 77~\% in
the $\Delta$ resonance and 50~\% in the second resonance region.

\subsubsection{SAID parameterization}

The SAID group parameterizes the world data in the form~\cite{Arn02}
\begin{equation}
T^\alpha (W,Q^2)= [t_{Born}^\alpha (W,Q^2) + A^\alpha ] \, [1 + i T_{\pi
N}^\alpha (W)] + B^\alpha \,T_{\pi N}^\alpha(W) + (C^\alpha+ i D^\alpha)[Im
T_{\pi N}^\alpha (W)-|T_{\pi N}^\alpha(W)|^2] \, . \label{eq:3.9}
\end{equation}
Since $T_{\pi N}= \mbox{sin} \delta_{\alpha} \; e^{i \delta_{\alpha}}$ below
two-pion threshold, the first (background) term on the r.h.s. of
Eq.~(\ref{eq:3.9}) is proportional to $\mbox{cos} \delta_{\alpha} \; e^{i
\delta_{\alpha}}$, the second (resonant) term is proportional to $\mbox{sin}
\delta_{\alpha} \; e^{i \delta_{\alpha}}$, and the last term vanishes in this
region of elastic scattering (except for small electromagnetic corrections).
The functions A to D are fitted to the data in the form of a polynomial
expansion in $W$ and $Q^2$.

\section{results}

In this section we compare the results of different approaches: the dynamic
models of Sato and Lee (S-L, Ref.~\cite{Sat96, Sat01}) and of the
Dubna-Mainz-Taipei collaboration (DMT, Ref.~\cite{Kam99, Kam01}), the effective
field theories of Pascalutsa and Vanderhaeghen (P-V, "$\delta$" expansion,
Ref.~\cite{Pas05}) and Gail and Hemmert (G-H, "$\epsilon$" expansion,
Refs.~\cite{Gai05, Gel99}), the phenomenological descriptions SAID~\cite{Arn02}
and MAID05~\cite{Dre99}, and the work of Hanstein based on dispersion relations
(DR, Ref.~\cite{Han96}). The basis of a comparison between theory and
experiment are the multipoles, which are complex functions of $W$ and $Q^2$.
Apart from corrections due to isospin symmetry breaking and electromagnetic
corrections, the Fermi-Watson theorem requires that all the $\Delta$ multipoles
$\mathcal{M}^{\alpha}$ carry the phase $\delta_{33}(W)$ in the energy range up
to the $\Delta$ region. As a consequence the real part of
$\mathcal{M}^{\alpha}$ vanishes at the $K$-matrix pole where $\delta_{33}
(W=M_{\Delta})=90^{\circ}$. The PDG~\cite{Yao06} lists $M_{\Delta}=(1232 \pm
1)$~MeV, but the description of the data may improve by changing the pole
position by several MeV. For example S-L  has the pole at 1230.3~MeV, and for
the listed SAID (SP06) solution the pole position is 1227.4~MeV for $\pi^0$
kinematics and 1232.8~MeV for $\pi^+$ kinematics. We further note that in the
following text all values refer to the $\Delta$ multipoles unless stated
otherwise, and we have therefore dropped the isospin index $I=3/2$ in our
notation.

\subsection{Delta(1232)  multipoles at the K-matrix pole }

\begin{figure}[h]
\includegraphics[height=.5\textheight]{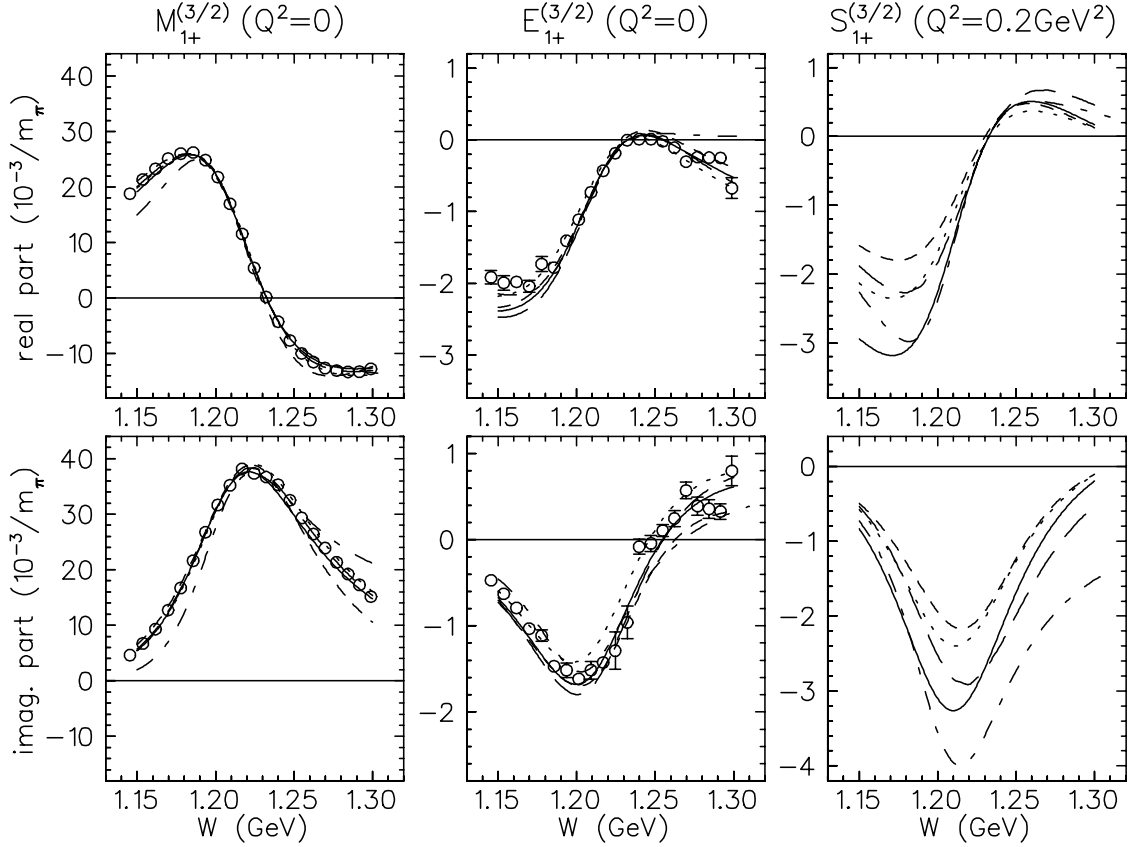}
\caption{The real and imaginary parts of the multipoles $M_{1+}^{(3/2)}$,
$E_{1+}^{(3/2)}$, and $S_{1+}^{(3/2)}$ as function of the c.m. energy W.  The
multipoles are in units of $10^{-3}/{\rm m}_{\pi^+}$. Solid line:
MAID05~\cite{Dre99}, long-dashed line: DMT~\cite{Kam99, Kam01}, short-dashed
line: S-L~\cite{Sat96, Sat01}, dashed-dotted line: P-V~\cite{Pas05}, dotted
line: SAID(SP06)~\cite{Arn02} for $n \pi^+$ kinematics. The data are
represented by the single energy solution of MAID05.} \label{fig:1}
\end{figure}

Figure~\ref{fig:1} shows the real and imaginary parts of the transverse $\Delta $
multipoles for $Q^2=0$ and the longitudinal multipole for $Q^2=0.2$~GeV$^2$ as function
of $W$. The imaginary parts at the K-matrix pole are listed in Table~\ref{tab:1} for a
more quantitative discussion. The data points in Fig.~\ref{fig:1} are obtained from the
single-energy analysis of MAID05. There is general agreement among the different
descriptions for the leading multipole $M_{1+}$, except that the curve S-L is slightly
shifted to the lower energies, whereas the curve P-V is lower at the left and higher at
the right shoulder of the resonance, which is probably due to the perturbative treatment
of the background in the $\delta $ expansion. Quite generally, the differences on both
shoulders of the resonance are mainly due to different treatments of the backgrounds, not
only of the non-resonant (3,3) background but also of other partial waves as
$E_{0+}^{(I)}, \, M_{1-}^{(I)}, \, E_{2-}^{(I)}$, and so on. Therefore different models
may see the data basis for the $\Delta$ multipoles somewhat different, not so much for
the leading magnetic transition but certainly for the much smaller electric and Coulomb
transitions.
\\

Seen in absolute values, the agreement is equally good for the multipole
$E_{1+}$, however the differences become more visible because of the smallness
of this multipole. The double zero in ${\mathcal Re} (E_{1+})$ is remarkable,
the first one is enforced by the Fermi-Watson theorem while the second one
occurs because the large and negative non-resonant background takes over both
below and above the resonance. The moderate differences for ${\mathcal Im}
(E_{1+})$ lead to large relative differences if this value is compared at the
$K$-matrix pole $W=M_{\Delta} \approx 1232$~MeV, as can be seen more
quantitatively in Table~\ref{tab:1}. Because $M_{1+}$ is pretty stable, it is
convenient to compare the ratios $R_{EM}= E_{1+}/ M_{1+}$ and $R_{SM}= S_{1+}/
M_{1+}$ at the $K$-matrix pole. In this context it is worth pointing out that
these ratios change rapidly as function of $W$: For example, within MAID and
for $Q^2=0$, $R_{EM}$ increases from $-7.5~\%$ at 1182~MeV and $-2.2~\%$ at the
K-matrix pole to $+2.6~\%$ at 1282~MeV. The low value $R_{EM}=-1.3~\%$ of SAID
is obtained with the solution SP06 and $\pi^0$ kinematics, earlier
fits~\cite{Arn02} yielded $R_{EM}=-1.79~\%$ at $Q^2=0$. Because the multipoles
are not yet available for the $\epsilon$ expansion of G-H~\cite{Gai05}, we have
calculated them from the complex form factors by means of Eq.~(\ref{eq:2.10}).
Because of the perturbative approach in that reference, the ratios $R_{EM}$ and
$R_{SM}$ become complex numbers indicating a strongly non-unitary result. The
ratios are therefore listed in Table \ref{tab:1} with the definition of the
authors, $R_{EM}={\mathcal Re} [E_{1+}/M_{1+}]$ etc., which of course in a
unitary theory is identical with our definition but neglects the huge imaginary
part in the non-unitary theory. More specifically, the phase at the K-matrix
pole should be $90^{\circ}$ for each multipole, whereas the multipoles of G-H
have the phases $83^{\circ} (M_{1+})$, $-170^{\circ} (E_{1+})$, and
$-146^{\circ} (S_{1+})$. Moreover, the neglected imaginary part of the ratio
$R_{EM}$ is more than three times the real part shown in Table~\ref{tab:1}. We
therefore suggest to fit the LECs to the behavior at the T-matrix pole and not
at the K-matrix pole (see below).

\newpage

%%%%%%%%%%%%%%%%% figure 1 %%%%%%%%%%%%%%%%%%%%
%%%%%%%%%%%%%%%%% figure 1 %%%%%%%%%%%%%%%%%%%%

%%%%%%%%%%%% table 1 %%%%%%%%%%
\begin{table}[h]
\begin{tabular}{c|cc|cc|cc|cc|cc|l}
\hline & \tablehead{2}{c}{c}{${\mbox{M}_{1+}}$ } &
\tablehead{2}{c}{c}{${\mbox{E}_{1+}}$ } &
\tablehead{2}{c}{c}{${\mbox{S}_{1+}}$} &
\tablehead{2}{c}{c}{$\mbox{R}_{EM}~[\%]$ } &
\tablehead{2}{c}{c}{$\mbox{R}_{SM}~[\%]$} &
\tablehead{1}{c}{c}{references} \\
\hline
$Q^2$ [GeV$^2$] & 0 & 0.2 & 0 & 0.2 & 0 & 0.2 &0 & 0.2 & 0 & 0.2 &\\
\hline
S-L\tablenote{pole at 1230.3~MeV} & 37.9 & 41.0 & -1.03 & -1.31 & -0.88 & -1.93 & -2.7 & -3.2 & -2.3 & -4.7 &\cite{Sat96, Sat01}\\
DMT & 37.2 & 39.7 & -0.88 & -1.13 & -1.70 & -2.59 & -2.4 & -2.8 & -4.6 & -6.5 &\cite{Kam99, Kam01}\\
P-V & 36.3 & 37.2 & -0.83 & -1.14 & -1.28 & -3.52 & -2.3 & -3.1 & -3.5 & -9.5 &\cite{Pas05}\\
G-H\tablenote{imaginary parts of multipoles and real parts of the ratios of
multipoles, as obtained from the complex form factors}
& 36.5 & 37.1 & -0.55 & -0.36 & -1.43 & -2.43 & -2.5 & -1.8 & -4.6 & -6.9 &\cite{Gai05}\\
SAID\tablenote{solution SP06, $p \pi^0$ kinematics} & 37.1 & 39.8 & -0.47
&-0.37& -1.54 &-2.04 & -1.3 & -0.9 & -4.2 & -5.1 &\cite{Arn02}\\
MAID & 36.0 & 39.0 & -0.78 & -0.79 & -2.38 & -2.58 & -2.2 & -2.0 & -6.6 & -6.6 &\cite{Dre99}\\
DR & 37.9 &  & -0.89 &  &  &  & -2.4 &  &  &  &\cite{Han96}\\
\hline
\end{tabular}
\caption{The imaginary parts of the multipoles $M_{1+}^{(3/2)}$,
$E_{1+}^{(3/2)}$, and $S_{1+}^{(3/2)}$ as well as the ratios $\mbox{R}_{EM}$
and $\mbox{R}_{SM}$ at the $K$-matrix pole for $Q^2$=0 and 0.2 GeV$^2$. The
multipoles are in units of $10^{-3}/{\rm m}_{\pi^+}$.} \label{tab:1}
\end{table}
%%%%%%%%%%%%%%%%% table 1 %%%%%%%%%%%%%%%%%%%%

\newpage

A view at the Coulomb multipole $S_{1+}$ in Fig.~\ref{fig:1} and Table~\ref{tab:1} shows
that the model predictions differ by up to a factor of two. In fact we have chosen
$Q^2=0.2$ GeV$^2$ in the figure, because the comparison at $Q^2=0$ looks even worse. Of
course, this multipole can never be measured at the real photon point, but the recent
experiments of the Bates/Mainz Collaboration at small values of $Q^2$ down to 0.05
GeV$^2$ will certainly lead to a convergence of the models, which partly differ because
of lower, now withdrawn or corrected data points near $Q^2=0.12$ GeV$^2$. Another (mild)
constraint is the pseudo-threshold relation, Eq.~({\ref{eq:2.7}}), requiring $S_{1+}$ to
vanish at $Q^2=-0.086$ GeV$^2$ with a curvature related to the slope of $E_{1+}$. In this
context we repeat that gauge invariance yields the relation
$L_{1+}(Q^2=0)=S_{1+}(Q^2=0)$, whereas the relation $E_{1+}=L_{1+}$ is only valid at
pseudothreshold.
\\

\begin{figure}[h]
\includegraphics[height=.25\textheight]{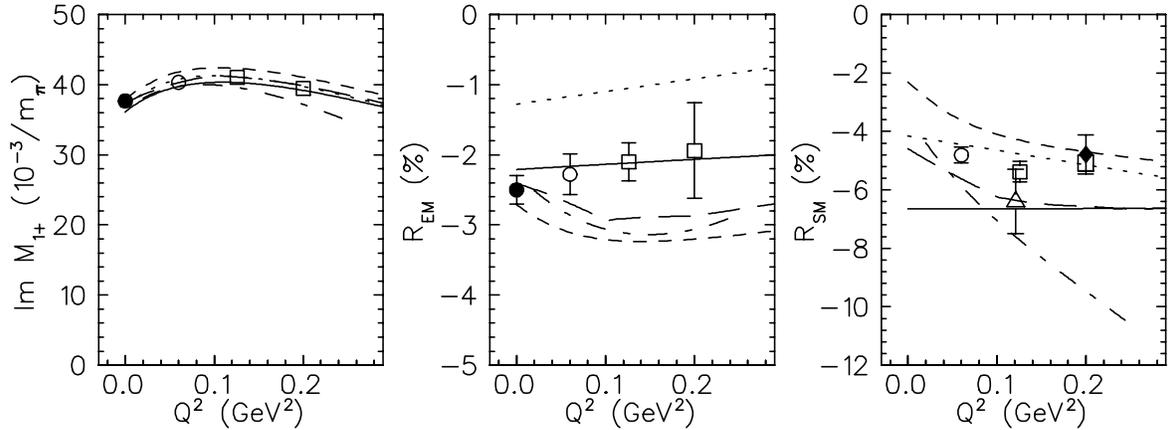}
\caption{The imaginary part of the multipole $M_{1+}^{(3/2)}$ and the ratios
$\mbox{R}_{EM}$ and $\mbox{R}_{SM}$ at the $K$-matrix pole as function of
$Q^2$. The multipole is given in units of $10^{-3}/{\rm m}_{\pi^+}$, the ratios
in $\%$. Solid line: MAID05~\cite{Dre99}, long-dashed line: DMT~\cite{Kam99,
Kam01}, short-dashed line: S-L~\cite{Sat96, Sat01}, dashed-dotted line: P-V
~\cite{Pas05}, dotted line: SAID (SP06)~\cite{Arn02} for $p \pi^0$ kinematics.
The experimental data are from Refs.~\cite{Bec97} (full circle), \cite{Sta06}
(open circle), \cite{Pos01} (open triangle), \cite{Spa05} (open squares), and
\cite{Els06} (full diamond).} \label{fig:2}
\end{figure}

The dependence of the multipoles on momentum transfer $Q^2$ is shown in Fig.
\ref{fig:2}. The results for ${\mathcal Im} (M_{1+})$ are reasonably well
reproduced by all authors. We note that the decrease towards $Q^2=0$ is
enforced by the proportionality of $M_{1+}$ with $k$ near pseudothreshold,
$Q^2=-0.086$ GeV$^2$. In the same limit, the ratio $R_{EM}$ approaches a
constant, whereas $R_{SM}$ has to vanish. We observe that the different
approaches yield quite different values for both $R_{EM}$ and $R_{SM}$, but we
repeat that also the data evaluation may depend on the backgrounds, which
differ to some extent from model to model.

%%%%%%%%%%%%%%%%% figure 2 %%%%%%%%%%%%%%%%%%%%
%%%%%%%%%%%%%%%%% figure 2 %%%%%%%%%%%%%%%%%%%%

\subsection{Delta(1232) multipoles at the T-matrix pole }

In this subsection we study the resonance structure of the different
approaches: the position of the resonance pole in the complex $W$ plane, the
size of the resonance contribution at the pole, and the angle by which the
non-resonant background rotates the phase of the resonance amplitude. Using the
described speed plot technique and averaging over the results of the groups
listed in Table \ref{tab:2}, we find from the multipole $M_{1+}$ and for
$Q^2=0$ the pole position at $W=M_R - i \Gamma_R /2$, with $M_R=(1212 \pm
4)$~MeV and $\Gamma_R = (101 \pm 10$)~MeV. The width obtained from the $\delta$
expansion~\cite{Pas05} is only $81$~MeV and well below the 1-$\sigma$ range. We
may attribute this small width to the perturbative treatment of the EFT, which
should of course improve by going to higher order in the expansion. If we
exclude this information from the average, the errors decrease by a factor of
about two. The obtained values for the pole position compare well with the
results from pion-nucleon scattering, $(1210 \pm 1)$~MeV for the real part and
($100 \pm 2$)~MeV for the width ~\cite{Yao06}. Quite similar results are
obtained from the 2 other multipoles and as function of $Q^2$. The pole
parameters $Z_{M1+}$ obtained by the different approaches are listed in Table
\ref{tab:2} for $Q^2=0$ and $0.2$ GeV$^2$. As average of the modulus we obtain
$r_{M1+}=(20.9 \pm 1.3)10^{-3}/ {\rm m}_{\pi+}$. The average phase of the
residue is $\Phi_{M1+}=(-26.4 \pm 11.0)^{\circ}$. This number is in the same
ballpark as the background phase given by Höhler~\cite{Hoe92}, $-24^{\circ}$,
and about half of the corresponding phase in pion-nucleon scattering,
$-47^{\circ}$~\cite{Yao06}. The results for the small quadrupole transition are
$r_{E1+}=(1.21 \pm 0.12)10^{-3}/ {\rm m}_{\pi+}$ and $\Phi_{E1+}=(-155 \pm
8)^{\circ}$. From this we obtain the ratio $r_{E1+}/r_{M1+} = (5.8 \pm 1.0)~\%$
or, if we exclude Ref.~\cite{Pas05} again, we have $r_{E1+}/r_{M1+} = (5.8 \pm
0.5)~\%$. The small deviation of less than 10~\% contrasts with the ratio at
the $K$-matrix pole, which is much more model-dependent (see
Table~\ref{tab:1}). These findings agree with earlier work~\cite{Han96, Wor99}
showing that the ratio $R_{EM}$ is more stable at the T-matrix pole than at the
K-matrix pole. Similar to the situation at the $K$-matrix pole, the deviations
for $S_{1+}$ are much larger, $r_{S1+}=(1.15 \pm 0.45)10^{-3}/ {\rm m}_{\pi+}$
and $\Phi_{S1+}=(-177 \pm 23)^{\circ}$.
\\

%%%%%%%%%%%% table 2 %%%%%%%%%%
\begin{table}[h]
\begin{tabular}{c|cc|cc|cc|l}
\hline & \tablehead{2}{c}{c}{$\mbox{Z}_{M1+} \; [10^{-3}/{\rm m}_{\pi+}]$} &
\tablehead{2}{c}{c}{$\mbox{R}_{EM}^{pole} \; ~[\%]$} &
\tablehead{2}{c}{c}{$\mbox{R}_{SM}^{pole} \; ~[\%]$} &
\tablehead{1}{c}{c}{references} \\
\hline
$Q^2$ [GeV$^2$] & 0 & 0.2 & 0 & 0.2 & 0 & 0.2 & \\
\hline
S-L\tablenote{$K$-matrix pole at 1230.3~MeV} & 21.9-4.7 i & 24.1-3.7 i &-4.3-3.3 i & -4.0-2.3 i & -3.2-2.7 i & -5.3-1.7 i & \cite{Sat96, Sat01}\\
DMT & 19.3-8.5 i & 21.0-8.3 i &-3.8-5.3 i & -3.9-3.4 i & -4.9-3.8 i & -7.6-2.1 i & \cite{Kam99, Kam01}\\
P-V & 12.9-13.2 i & 14.8-13.1 i & -3.0-4.5 i & -3.0-4.3 i & -3.8-3.2 i & -9.6-1.9 i & \cite{Pas05}\\
SAID\tablenote{solution SP06, $n \pi^+$ kinematics} & 18.6-9.7 i & 20.7-9.0 i &-2.8-4.8 i & -2.0-3.4 i & -3.5-0.2 i& -5.8-2.7 i & \cite{Arn02}\\
MAID& 19.9-7.9 i & 22.5-6.9 i &-3.7-4.4 i & -3.1-3.1 i & -7.5-4.5 i & -7.5-3.2 i & \cite{Dre99}\\
DR & 18.8-9.8 i&  &-3.5-4.6 i & &  & & \cite{Han96}\\
\hline
\end{tabular}
\caption{The strength of the multipole $M_{1+}^{(3/2)}$, as described by the
resonance pole parameter $Z_{M1+}$ according to Eq.~(\ref{eq:2.4b}), and the
ratios $R_{EM}^{pole}$ and $R_{SM}^{pole}$ at the $T$-matrix pole for $Q^2$=0
and 0.2 GeV$^2$.} \label{tab:2}
\end{table}
%%%%%%%%%%%% table 2 %%%%%%%%%%

Finally, we compare these results with the $\epsilon$ expansion of Gellas
\emph{et al.}~\cite{Gel99} who calculated the complex form factors for the
decay $\Delta \rightarrow N + \gamma$ in the framework of a chiral EFT. Since
this work dresses the $\Delta$ by pion-nucleon and pion-$\Delta$ loops but
neglects the possibility of producing a non-resonant pion-nucleon system at the
energy of the $\Delta$, it should be related to the situation at the T-matrix
pole. And indeed, one obtains $r_{E1+}/r_{M1+} = (5.8 \pm 0.2)~\%$ from this
reference, in agreement with the result presented above. However, the angles
$\Phi_{E1+}$ and $\Phi_{M1+}$ differ from our findings, in particular
$\Phi_{E1+}$ is rotated by about $100^{\circ}$ against our solution.
Qualitatively similar results are obtained in the recent work of
Ref.~\cite{Gai05} except that the ratio $r_{E1+}/r_{M1+}$ increases to
$8.4~\%$.

\subsection{Quark bag vs. pion cloud physics}

\begin{figure}[h]
\includegraphics[height=.27\textheight]{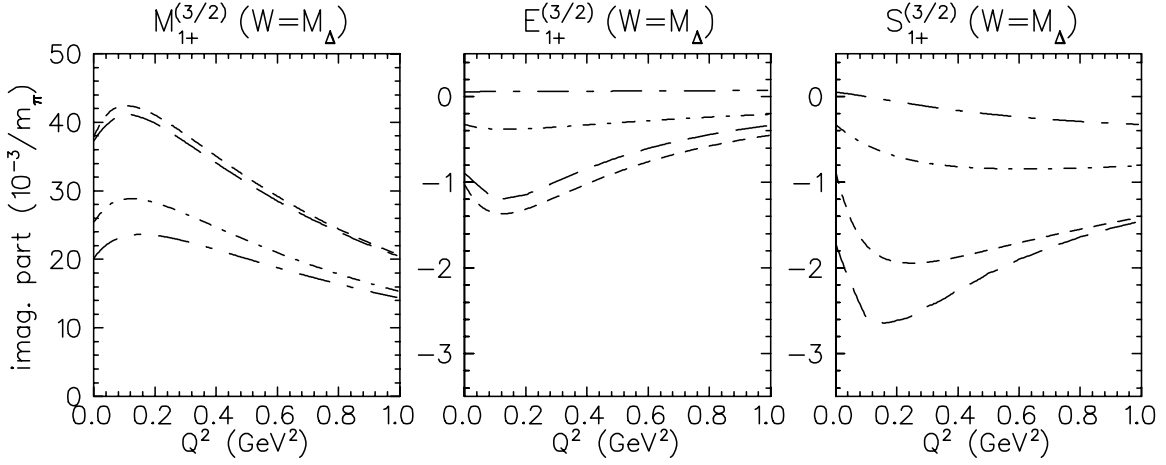}
\caption{The imaginary parts of the multipoles $M_{1+}^{(3/2)}$,
$E_{1+}^{(3/2)}$, and $S_{1+}^{(3/2)}$ at the K-matrix pole as function of $Q^2$. The
multipoles are given in units of $10^{-3}/{\rm m}_{\pi^+}$. Short-dashed line:
S-L~\cite{Sat96, Sat01}, short-dashed-dotted line: S-L with the bare $\Delta$
contribution only, long-dashed line: DMT~\cite{Kam99, Kam01}, long-dashed-dotted line:
DMT with the bare $\Delta$ contribution only.} \label{fig:3}
\end{figure}

Figure~\ref{fig:3} compares the full multipoles at the $K$-matrix pole with the
corresponding results for the "bare" $\Delta$, and in Table~\ref{tab:3} we list
the values of the three $\Delta$ multipoles as well as the ratios $R_{EM}$ and
$R_{SM}$ for $Q^2=0$ and $Q^2=0.2$~GeV$^2$. The multipoles are plotted as
function of $Q^2$ in the range from the real photon point to a virtuality of 1
GeV$^2$. The pion rescattering or "pion cloud" contribution is quite
substantial, as can be seen from the region between the respective dashed and
dashed-dotted lines in Fig.~\ref{fig:3}. In the case of the magnetic dipole
transition at $Q^2=0.2$ GeV$^2$, the "bare" amplitude $M_{1+}$ contributes only
$70~\%$ (S-L) or $60~ \%$ (DMT) to the full amplitude, that is, the pion cloud
yields about one third of the transition strength. This explains why quark
models chronically underestimate the $M1$ transition by this amount. The figure
also shows that the pionic contribution drops from about 1/3 of the total
amplitude at the maximum to about 1/4 at $Q^2=1$ GeV$^2$. In other words, the
long-range physics drops out faster with increasing resolution due a typical
radius of, say, 1 fm for the pion cloud, whereas the short-distance physics
with a typical quark-bag of radius 0.5 fm decreases slower as function of
$Q^2$. The two models differ more strongly for the electric and Coulomb
multipoles. In particular the bare $\Delta$ contribution to $E_{1+}$ is
negligible for the DMT model, whereas the S-L model obtains a $30~\%$
contribution from the bare $\Delta$. Since both models describe the data
reasonably well, we may ask what is the physical reason for this difference.
\\

%%%%%%%%%%%% figure 3 %%%%%%%%%%

%%%%%%%%%%%% figure 3 %%%%%%%%%%

As we have seen before, the decomposition of "background" vs. "resonant"
contributions is handled differently in the two models. In the DMT model, the
background contribution refers to the non-resonant production of a pion ("the
photon interacts initially with the pion cloud"), which after its production is
re-scattered by the full hadronic interaction corresponding to the phase
$\delta_{33}$. We can discuss this aspect more quantitatively by speed-plotting
the two contributions for the DMT model. At $Q^2=0$ we find the tiny ratio
$R_{EM}^{pole}=(0.2+0.2 i)~\%$ for the bare $\Delta$, which is to be compared
with the value $R_{EM}^{pole}=(-3.8-5.3 i)~\%$ for the full DMT model given in
Table~\ref{tab:2}. Therefore the "background" contribution of the DMT model is
essentially responsible for the residues of the small multipoles $E_{1+}$ and
$S_{1+}$ at the T-matrix pole. In contrast, the slowly changing phase
$\delta_b$ appearing in the background term of the S-L model does not give rise
to a pole with the properties of the $\Delta$, and as a consequence the
residues of the electric and Coulomb strengths can not be provided by the
background term of the S-L model. And indeed, the large effect of non-resonant
pion production followed by $\Delta$ excitation via rescattering, as observed
in the DMT model, is now absorbed in the renormalization of the $\gamma + N
\rightarrow \Delta$ vertex of the S-L model. Of course, the different
definitions do not explain why the "bare $\Delta$" contribution differs so much
in the two models. However, the huge effect of $\Delta$ production by
rescattering may possibly give us some hint. Whereas the on-shell contribution
of the rescattering term should be essentially model-independent, the principal
value integral contains off-shell kinematics and therefore is very sensitive to
model assumptions. In order to fully understand the issue, it would be
interesting to compare the off-shell properties of the hadronic interaction in
the two models and the resulting consequences for the bare $\Delta$ vs. pion
cloud contributions.
\\

%%%%%%%%%%%% table 3 %%%%%%%%%%
\begin{table}[h]
\begin{tabular}{c|cc|cc|cc|cc|cc|l}
\hline & \tablehead{2}{c}{c}{${\mbox{M}_{1+}}$ } &
\tablehead{2}{c}{c}{${\mbox{E}_{1+}}$ } & \tablehead{2}{c}{c}{${\mbox{S}_{1+}}$
}& \tablehead{2}{c}{c}{$\mbox{R}_{EM}~[\%]$ } &
\tablehead{2}{c}{c}{$\mbox{R}_{SM}~[\%]$} &\tablehead{1}{c}{c}{references} \\
\hline
$Q^2$ [GeV$^2$] & 0 & 0.2 & 0 & 0.2 & 0 & 0.2 &0 & 0.2 & 0 & 0.2 &\\
\hline
S-L full\tablenote{pole at 1230.3~MeV}& 37.9 & 41.0 & -1.03 & -1.31 & -0.88 & -1.93 & -2.7 & -3.2 & -2.3 & -4.7 & \cite{Sat96, Sat01}\\
bare& 25.4 & 28.3 & -0.33 &-0.38 & -0.32 & -0.70 & -1.3 & -1.3 & -1.3 & -2.5 & \\
\hline
DMT full& 37.1 & 39.7 & -0.87 & -1.12 & -1.51 & -2.64 & -2.3 & -2.8 & -4.1 & -6.7 &  \cite{Kam99, Kam01}\\
bare& 20.2 & 23.5 & 0.05 & 0.06 & 0.05 & -0.06 & 0.3 & 0.3 & 0.3 & -0.3 &\\
\hline
G-H full\tablenote{imaginary parts of multipoles
as obtained from the real part of the complex form factors}& 36.5 & 37.1 & -0.55 & -0.36 & -1.43 & -2.43 & -2.5 & -1.8 & -4.6 & -6.9 &  \cite{Gai05}\\
bare& 49.3 &  & 1.34 &  & 0.47 &  &  &  &  &  &\\
\hline
\end{tabular}
\caption{The imaginary parts of the multipoles $M_{1+}^{(3/2)}$,
$E_{1+}^{(3/2)}$, and $S_{1+}^{(3/2)}$ as well as the ratios $\mbox{R}_{EM}$ and
$\mbox{R}_{SM}$ at the $K$-matrix pole for $Q^2$=0 and 0.2 GeV$^2$. The multipoles are in
units of $10^{-3}/{\rm m}_{\pi^+}$. The results for the full 33-partial waves are listed
in the upper row for each reference, the values for the bare $\Delta$ resonance are in
the lower row.} \label{tab:3}
\end{table}
%%%%%%%%%%%% table 3 %%%%%%%%%%

Let us now compare the results of the dynamic models with the $\epsilon$
expansion of Ref.~\cite{Gai05}, which splits the short-distance physics and the
pion-cloud contribution at a scale of 1 GeV. The entry G-H in Table \ref{tab:3}
shows that the pion cloud yields a negative contribution to $M_{1+}$, at
variance with the dynamic models, which predict a positive contribution to the
magnetic transition. As a consequence the fit to the data requires a "bare"
$\Delta$ form factor much above the data. This contradicts the common belief
that the constituent quark models can be brought to agree with the data by
including the pionic degrees of freedom. In the case of $E_{1+}$, the pionic
contribution of the $\epsilon$ expansion is 2-3 times larger than predicted by
the dynamic models, which requires a large cancelation with the short-distance
physics. The same cancelation occurs also for $S_{1+}$, but to a lesser degree.
In conclusion all approaches agree that the transverse electric and Coulomb
N-$\Delta$ transitions are essentially determined by the pion cloud. However,
there remain serious numerical contradictions among the different descriptions.
In order to clarify this issue it will be important to compare at the level of
the multipoles, because only these are directly related to the experiment.
Since the multipoles fulfill the Fermi-Watson theorem in the $\Delta$ region,
their ratios are real numbers, which is not the case for the form factors of
Refs.~\cite{Gel99,Gai05}. Therefore we suspect that some of the differences are
due to the (partial) neglect of final-state interactions, which are expected to
affect the multipoles quite differently.
\\

Another point requiring some common studies is the dependence of the "bare vs.
dressed $\Delta$" issue on the cut-off of the loop integrals or the scale of
the separation. Of course, we are aware of the fact that a "pion cloud" or
"bare" and "dressed" $\Delta$" resonances are not observables, but still the
physics behind the mere $S$-matrix elements remains intriguing. However, the
picture we have in mind about the nucleon - an interior region of valence
quarks, glue, and quark-antiquark pairs surrounded by a pion cloud - does only
make sense if different approaches agree at least semi-quantitatively under the
same conditions, for example, at the same scale of regularization.

\begin{theacknowledgments}
We are grateful for helpful suggestions and numerical input to R.A. Arndt, G.C.
Gail, S.S. Kamalov, T.-S.H. Lee and T. Sato, V. Pascalutsa and M.
Vanderhaeghen. This work was supported by the Deutsche Forschungsgemeinschaft
(SFB 443).
\end{theacknowledgments}

\end{document}